\documentclass[preprint]{aastex61}
\usepackage{natbib}
\usepackage{graphicx}
\usepackage{amsmath}
\usepackage{amsfonts}
\usepackage{amssymb} 
\usepackage[]{units}
\usepackage{nicefrac}
\usepackage{longtable}
\usepackage{bm}
\usepackage{mathtools}
\DeclareMathAlphabet{\mathcal}{OMS}{cmsy}{m}{n}
\SetMathAlphabet{\mathcal}{bold}{OMS}{cmsy}{b}{n}

\newcommand{\e}[1]{\ensuremath{\times 10^{#1}}}

\newcommand{\set}[1]{\bm{#1}}
\newcommand{\starlabel}{\ell}
\newcommand{\starlabelvec}{\set{\starlabel}}
\newcommand{\mean}[1]{\overline{#1}}
\newcommand{\given}{\,|\,}

\newcommand{\scatter}{s_{\lambda}}

\newcommand{\chisq}{\ensuremath{\chi^2}}
\newcommand{\teff}{\ensuremath{T_{\text{eff}}}}

\newcommand{\vsini}{\ensuremath{v\sin{i}}}
\newcommand{\teffe}{\ensuremath{T_{\text{eff}}}~}

\newcommand{\vsinie}{\ensuremath{v\sin{i}}~}
\newcommand{\feh}{\ensuremath{[\text{Fe}/\text{H}]}}
\newcommand{\fehe}{\ensuremath{[\text{Fe}/\text{H}]}~}
\newcommand{\met}{\ensuremath{[\text{M}/\text{H}]}}

\newcommand{\logg}{\ensuremath{\log{g}}}
\newcommand{\logge}{\ensuremath{\log{g}}~}

\newcommand{\kms}{\ensuremath{\text{km}\,\text{s}^{-1}}}
\newcommand{\df}{\ensuremath{f^\prime}}
\newcommand{\bigo}{\ensuremath{\mathcal{O}}}

\newcommand{\apogee}{\textsl{APOGEE}}

\newcommand{\nt}[1]{#1}
\newcommand{\nnt}[1]{#1}

\graphicspath{ {./}{./Figures/} }

\shorttitle{A Data-Driven Technique for Measuring Stellar Rotation}
\shortauthors{Gilhool \& Blake}

\begin{document}

\title{A Data-Driven Technique for Measuring Stellar Rotation}

\author[0000-0002-3230-3052]{Steven H. Gilhool}
\affiliation{Department of Physics and Astronomy, University of Pennsylvania, 209 S. 33rd Street, Philadelphia, PA 19104}
\author[0000-0002-6096-1749]{Cullen H. Blake}
\affiliation{Department of Physics and Astronomy, University of Pennsylvania, 209 S. 33rd Street, Philadelphia, PA 19104}

\begin{abstract}
Measuring stellar rotational velocities is a powerful way to probe the many astrophysical phenomena that drive, or are driven by, the evolution of stellar angular momentum. In this paper, we present a novel data-driven approach to measuring the projected rotational velocity, \vsini. Rather than directly measuring the broadening of spectral lines, we leverage the large information content of high-resolution spectral data to empirically estimate \vsini.  We adapt the framework laid down by The Cannon (\citealt{Ness15}), which trains a generative model of the stellar flux as a function of wavelength using high-fidelity reference data, and can then produce estimates of stellar parameters and abundances for other stars directly from their spectra. Instead of modeling the flux as a function of wavelength, however, we model the first derivative of the spectra, as \nnt{we expect} the slopes of spectral lines \nnt{to} change as a function of \vsini. This technique is computationally efficient and provides a means of rapidly estimating \vsinie for large numbers of stars in spectroscopic survey data. We analyze SDSS APOGEE spectra, constructing a model informed by high-fidelity stellar parameter estimates derived from high-resolution California Kepler Survey spectra of the same stars.  We use the model to estimate \vsinie up to $\unit[15]{\kms}$ for $27,000$ APOGEE spectra, in fractions of a second per spectrum. Our estimates agree with the APOGEE \vsinie estimates to within $\unit[1.2]{\kms}$.
\end{abstract}


%\maketitle

\section{Introduction}

Measuring stellar rotation is a powerful way to probe many astrophysical phenomena. The magnitude of stellar angular momentum and the rate at which it is transferred to the surrounding medium are closely tied to many astrophysical processes such as the outflow of stellar material; the magnetic dynamo mechanism; and the strength, morphology and lifetime of the stellar magnetic field. Due to conservation of angular momentum, the stellar angular momentum increases as the star collapses onto the main sequence. The subsequent evolution of stellar angular momentum is thought to be governed by interactions between a star's magnetic field, photosphere, and the ambient environment.

The average rotation rate of stars is known to vary across the main sequence, reflecting important differences in stellar structure and magnetic field properties. In the middle of the main sequence, roughly between spectral types mid-F and mid-K, stars are found to be rotating slowly. Sun-like field stars typically rotate at velocities of $v_{\textrm{rot}} \lesssim \unit[5]{\kms}$. \nt{These stars primarily spin down through magnetic braking. The magnetic field, generated at the interface between the star's radiative core and its convective envelope, interacts with the stellar wind and surrounding medium, and causes the star to lose angular momentum at a rapid rate.} At either end of the main sequence, however, there is a sharp increase in the average rotation rate \citep{Sta86}. \nt{Stars more massive than $\sim \unit[1.3]{M_\sun}$ ($>$ F5) lack an outer convective envelope, and therefore do not produce surface magnetic fields of sufficient extent to efficiently brake the star \citep{Schatz62}. The mean rotation rate for massive stars is on the order of $\unit[150]{\kms}$.} Meanwhile, at the bottom of the main sequence, late M dwarfs and brown dwarfs are thought to rotate rapidly due to being \textit{fully} convective. \nt{Lacking the radiative core, these stars are thought to produce topologically distinct magnetic fields, which similarly may not be able to efficiently dissipate angular momentum (\citealt{Stas11}, \citealt{Bou13}, \citealt{Houd17}).}

Several methods can be used to measure stellar rotation, with different methods applicable in different areas of parameter space. For spotted stars that may exhibit photometric variability related to rotation, photometric rotation periods can be inferred. For example, given the excellent photometric precision of the Kepler satellite, rotation periods for thousands of stars have been measured (e.g., \citealt{mcquillan2014}). This method is,  depending on the time baseline and sampling of the photometry, sensitive to rotation on a wide range of timescales, but is not effective for stars that lack spots. At the same time, spot evolution on timescales that are short compared to the rotation period may lead to ambiguity in the estimated rotation periods. 

The projected rotational velocity, \vsini, can be inferred from the analysis of line broadening in high-resolution spectra. A broadened line profile is equivalent to the convolution of the intrinsic profile and a rotational broadening kernel. In the Fourier domain, this convolution becomes a product, and so the Fourier transform of the broadening kernel can be isolated. The Fourier transform of the rotational broadening kernel has zeros at frequencies inversely proportional to \vsini~(e.g., \citealt{Dra90}, \citealt{Diaz11}). Estimating \vsinie from the Fourier transform of line profiles requires isolated lines and high-resolution, high signal-to-noise spectra. Additionally, this approach is best-suited for use with rapid rotators ($\vsini \gtrsim \unit[30]{\kms}$) as the first zero of the Fourier transform of the broadening kernel can be obscured by high-frequency Fourier noise at lower \vsinie \citep{Bou13}. 

Alternatively, \vsinie can be measured using the cross-correlation method (e.g., \citealt{Del98}, \citealt{Rei12}, \citealt{Houd15}).  Here, a star known to be rotating slowly is used as a template. The template star is broadened at various values of \vsini, and the broadened copies are cross-correlated with the unbroadened template. The widths of the resulting cross-correlation function peaks provide a calibrated measure of \vsini. Observed spectra are then cross-correlated with the template, and their \vsinie is inferred from the width of the cross-correlation peak. This method requires template stars (known to be rotating slowly) of the same spectral type as the observed stars.

Finally, a straightforward template-fitting technique (e.g., \citealt{Jen09}, \citealt{Pass16}, \citealt{Gil18}, \citealt{Raj18}) may be used, in which \vsinie is determined through the direct, pixel-by-pixel fitting of theoretical template spectra to observed spectra. Typically, a suite of stellar models is fit to the observed spectrum. The rotational broadening is simulated by the convolution of the templates with a broadening kernel, and the best-fit model yields the \vsinie measurement. This method is limited by its dependence on theoretical spectra, which may introduce (especially in the case of M dwarfs) biases related to systematic differences between stellar lines and lines in the theoretical templates.

In this paper, we present a novel, data-driven approach to measuring \vsini. Leveraging the large information content of high-resolution stellar spectra, we use regression techniques to estimate \vsinie empirically. \nt{Our approach is a modification of The Cannon (\citealt{Ness15}), a technique that trains a generative model of the stellar flux as a function of wavelength using high-fidelity reference data, and then produces estimates of stellar parameters and abundances for other stars directly from their spectra. Such a technique requires that the data set includes objects with high-fidelity measurements of \vsinie and other stellar parameters, spanning the parameter space of the full data set. We analyzed SDSS APOGEE spectra, constructing a model informed by high-fidelity stellar parameter estimates derived from high-resolution California Kepler Survey spectra of the same stars. The data are described in Section 2.}

\nt{The major change that we introduce to The Cannon framework is that, rather than modeling the \textit{flux} as a function of wavelength, we model the \textit{first derivative} of the flux, because we expect the slopes of spectral lines to depend strongly on \vsini. We describe the technique in detail in Section 3.}

\nt{While this approach to \vsinie estimation has its limitations, it is computationally efficient and provides a means of rapidly estimating \vsinie for large numbers of stars in spectroscopic survey data. We estimate \vsinie up to $\unit[15]{\kms}$ for $27,000$ APOGEE spectra, in fractions of a second per spectrum. Our estimates agree with the APOGEE pipeline \vsinie estimates to within $\unit[1.2]{\kms}$. Our results are discussed in Section 4. Finally, we discuss the advantages and limitations of our approach, outline some possible improvements, and summarize our findings in Section 5.}

% The primary advantage is that it is fast, and can be applied quickly to many thousands of stars in survey data. 

%In Section 2, we discuss the data that we used in this analysis.  In Section 3, we describe the technique.  In Section 4, we demonstrate our results. In Section 5, .

\section{Data Used in This Analysis}
\label{sec:data}

\subsection{Spectral Data from APOGEE}
We measure \vsinie using near-infrared spectra from Data Release 14 of the Sloan Digital Sky Survey (SDSS; \citealt{DR14}). The SDSS Apache Point Observatory Galactic Evolution Experiment (APOGEE) has produced near infrared spectra and associated data products, including estimates of stellar parameters and chemical abundances, for over 250,000 giant and dwarf stars. The APOGEE instrument (\citealt{Wil10}, \citealt{Skr15}) is a cryogenic, multiplexed near-infrared spectrograph with resolving power of $R\sim22,500$. Its three H-band detectors span a wavelength range of $\lambda = \unit[1.514]{\micron} - \unit[1.696]{\micron}$. The three chips are referred to respectively as `blue' ($\lambda = \unit[1.52]{\micron} - \unit[1.58]{\micron}$), `green' ($\lambda = \unit[1.59]{\micron} - \unit[1.64]{\micron}$), and `red' ($\lambda = \unit[1.65]{\micron} - \unit[1.69]{\micron}$). 

The main APOGEE pipeline reduces observations, removes telluric absorption and sky emission lines, and combines individual visit spectra from multiple epochs into single, high signal-to-noise (typically $> 100$) coadded spectra in the rest frame of the star. These apStar files contain the coadded and individual visit spectra for each star, along with important header information \citep{Nid15}. Additionally, APOGEE spectra are processed by the APOGEE Stellar Parameter and Chemical Abundance Pipeline (ASPCAP; \citealt{Gar16}), which produces psuedo-continuum normalized spectra, stellar parameter estimates and estimates of up to $15$ individual chemical elements. \nt{We applied the techniques described here to the apStar spectra. We used the ASPCAP parameter estimates to select the survey sample (see Section \ref{subsec:survey_sample}) from the full APOGEE catalog, and for later testing our results.} We restricted our sample to stars that were observed by the SDSS main $\unit[2.5]{m}$ telescope \citep{Gunn06}, and that have spectra with average signal-to-noise $> 100$ per pixel.

\subsection{Reference Data from the California Kepler Survey}
The data-driven technique described here depends on \nt{having} reference objects with high-fidelity parameter measurements, preferably derived from spectra of higher resolution than that of APOGEE.  For this purpose, we used \nt{stellar parameter estimates} from the California Kepler Survey (CKS), a project designed to precisely measure the properties of Kepler planets' host stars \nt{(mainly F-G-K dwarfs)}. The CKS data set consists of $1305$ high-resolution ($R = 60,000$) optical ($3,640$\,\AA\,$ \le \lambda \le 7,990$\,\AA) spectra from the Keck HIRES instrument (see \citealt{CKSI} and \citealt{CKSII}). \nt{While we performed early tests using the CKS spectra, the analysis presented here makes use of the CKS parameter estimates only.}

  \nt{The CKS team measured stellar parameters} using two different spectroscopic pipelines - SpecMatch \citep{Pet15} and SME@XSEDE \citep{Val96}.  In most cases, the CKS determinations of \teff, \feh, and \logge are the arithmetic means of the values output by the two pipelines.  The internal precision of CKS \teffe and \fehe measurements \nt{are} quoted at $\sigma_{\textrm{Teff}} = \unit[60]{K}$ and $\sigma_{\feh} = \unit[0.04]{dex}$\nt{, respectively}. Adding systematic uncertainties based upon comparison with other measurement techniques, the quoted total uncertainties are $\sigma_{\textrm{Teff}} = \unit[117]{K}$ and $\sigma_{\feh} = \unit[0.07]{dex}$. The total uncertainty in \logge is $\sigma_{\logg} = \unit[0.1]{dex}$ \citep{CKSI}. The CKS \vsinie values were determined solely by the SpecMatch pipeline. Based on a comparison between SpecMatch \vsinie and Rossiter-McLaughlin measurements \citep{Alb12} for a subset of the CKS data, the uncertainty in \vsinie was determined to be $\sigma_{\vsini} = \unit[1]{\kms}$ for $\vsini \ge \unit[1]{\kms}$. Stars with SpecMatch $\vsini < \unit[1]{\kms}$ are to be considered non-detections with upper-limits of $\vsini < \unit[2]{\kms}$ \citep{Pet15}. \nt{Out of the $1305$ CKS targets, $362$ were also observed by the APOGEE survey. The stars observed by both surveys formed the basis for our training sample.}

\subsection{Training Sample}
Our training sample is a subset of the $362$ stars with both APOGEE \nt{spectra and CKS parameter estimates}. We made quality cuts on the APOGEE spectra by requiring average signal-to-noise $\ge 100$ per pixel, and removing spectra with the \texttt{ASPCAP\_FLAG} set to \texttt{STAR\_BAD}, which indicates a critical failure in the ASPCAP pipeline. Using the CKS data, we \nt{further restricted} the sample to stars that are likely main sequence F-G-K dwarfs by requiring $\logg_{\textrm{CKS}} \ge 3.9$.  Finally, we restricted the range of \vsini, in order to have a sufficient density of training data spanning the entire range. Only seven of the APOGEE/CKS stars have $\vsini > \unit[14]{\kms}$, so we removed them. At the \nt{other extreme, $34$ stars had $\vsini_{\textrm{CKS}} < \unit[1]{\kms}$. We dealt with these non-detections} by creating two copies of the training sample. In the first, we used only the stars with detectably measured \vsinie ($\unit[1]{\kms} \le \vsini_{\textrm{CKS}} \le \unit[13.8]{\kms}$). \nt{This sample contained $236$ stars.} In the second, we \nt{incorporated} the non-detections \nt{as synthetic fast rotators}. \nt{That is,} we treated each of the non-detections as a non-rotating star, and broadened their spectra with uniform random values of \vsinie between $10$ and $\unit[15]{\kms}$, in order to supplement the number of training stars with $\vsini \ge \unit[10]{\kms}$ where the data was relatively sparse. This training sample contained $270$ spectra. We determined the complexity of the model by performing cross-validation tests on \nt{both of} these training samples. The model selection is described in more detail in Section \ref{subsec:model_selection}. Because the inclusion of artificial rotators significantly improved our results at $\vsini \ge \unit[10]{\kms}$, we adopted the latter set of $270$ APOGEE/CKS stars as our final training sample.

\begin{figure}[ht] %Training Data
    \includegraphics[width=1\textwidth]{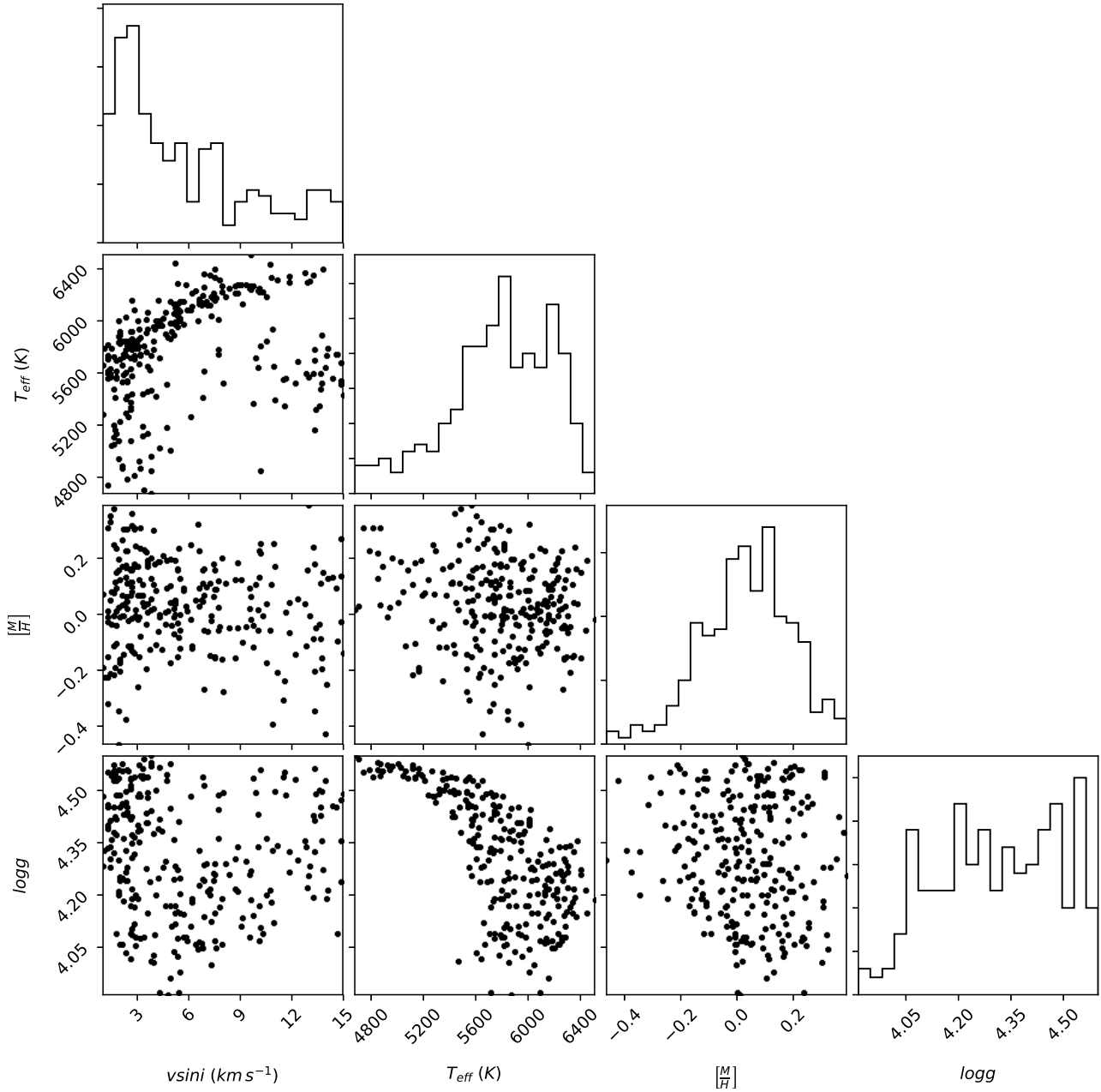}
    \caption{ Illustration of the CKS training data parameter space. The scatter plots show the training labels in several projections. The histograms show the distribution of training data for each label. }
    \label{fig:training_space}
\end{figure}

As measured by CKS, the training sample spans $\unit[4675]{K} \le \teff \le \unit[6508]{K}$, $\unit[-0.46]{} \le \feh \le 0.38$, and $\unit[3.9]{} \le \logg \le \unit[4.6]{}$. \nt{The parameter space spanned by the CKS reference labels is described in Figure \ref{fig:training_space}. The data from the final training set are plotted in various projections, with histograms showing the distributions of the training data for each label individually. Even with the addition of synthetic rotators, the training data are mostly slow rotators.} \nt{Though} the ASPCAP measurements of the \nt{training sample stars are not used in this analysis, we note that they are in close agreement with the CKS measurements. The ASPCAP values} range from $\unit[4660]{K} \le \teff \le \unit[6448]{K}$, and $\unit[-0.55]{} \le \feh \le 0.47$. None of the stars in the training sample have \logge measurements reported by ASPCAP. The ASPCAP headers do, however, contain \textit{uncalibrated} \logge estimates in the FPARAM vector. The \logge are not reported because ASPCAP does not have a \logge calibration for dwarf stars. The APOGEE website\footnote{\url{http://www.sdss.org/dr14/irspec/parameters/}} notes that ASPCAP \logge estimates are typically underestimated for dwarfs, but the FPARAM values happen to be in excellent agreement with the CKS \logg. \nt{We did, however, make use of the ASPCAP measurements (including the uncalibrated \logg) in selecting an appropriate survey sample out of the full APOGEE data set. }  

%% Describe bootstrap sample, if used
\subsection{Survey Sample}
\label{subsec:survey_sample}
\nt{The survey sample is the set of APOGEE stars similar to those in the training sample, but without CKS measurements. We selected APOGEE stars that were fit with the appropriate grids for F-G-K dwarfs (\texttt{ASPCAP\_CLASS} $=$ `GKd' or `Fd'), again with signal-to-noise $ > 100$, and in the same range of \teff, \feh, \logg, and \vsinie as spanned by the training set.} The exact ranges of the ASPCAP measurements of \nt{survey} sample stars are $\unit[4660]{K} \le \teff \le \unit[6439]{K}$, $\unit[-0.48]{} \le \feh \le 0.37$, and $\unit[3.9]{} \le \logg \le \unit[4.6]{}$. The \nt{survey} sample consists of precisely $27,000$ stars.

\nt{The aim of this method is to propagate the reference \vsinie measurements from the training sample to the stars in the survey sample. We now turn to describing, in detail, how we train the model, and how we use it to estimate \vsini.}

\section{Empirical Data-driven Technique}
\label{sec:method}

As the basis of our data-driven approach, we adopted the statistical framework of The Cannon (\citealt{Ness15}, \citealt{Cas16}).  The Cannon estimates stellar parameters by training a generative model, which describes the flux at each rest wavelength as a probability distribution in terms of the reference stellar parameters, hereafter referred to as stellar labels. The model can then be used to estimate labels for unlabeled \nt{survey} data\nt{, by determining the labels that most nearly generate the observed spectrum.}

% LIMITATION: Each pixel is independent
A key feature of The Cannon is that the model is trained at each wavelength independently. This rests on the assumption that stars with the same labels will have similar-looking spectra, and that the flux at each pixel will generally change smoothly as a function of the stellar labels. \citet{Ness15} showed that this assumption was warranted in the case of the three most important fundamental parameters, \teff, \feh, and \logg, which the Cannon recovers at higher precision than the reference ASPCAP data. 

Rotation, however, broadens and blends spectral lines, redistributing flux along the spectral dimension. Given two stars with identical \teff, \feh, and \logg, but very different rotation rates, the spectra will not look similar in terms of flux. As the Cannon sample was primarily composed of giant stars, rotational broadening was negligible in that analysis. In fact, the few stars that were flagged as rapidly rotating by the ASPCAP pipeline were excluded from the analysis. Our generative model instead describes the \textit{slope} of the spectrum at a given wavelength, rather than the flux. 

\nt{While differentiating the spectrum does not add information to the problem, and indeed, complicates the noise model by introducing correlation between pixels, we chose to work in flux-slope space for a few reasons.  The first is precisely \textit{because} of the introduction of correlation between pixels. \citet{Ness15} showed that it was reasonable to ignore the (actual) correlation between pixels when inferring \teff, \feh, and \logg, but because rotation increases the correlation of the flux between pixels, we felt that the assumption of independence between pixels becomes more problematic as \vsinie increases. Instead, we make the assumption that we can treat the flux-slope at each pixel as independent. The second is that it is possible that there is less correlation between the labels (\teff ~and \feh, for example) when working in flux-slope space. Finally, we expect the slope to be less affected by errors in continuum normalization. \citet{Ness15} and \citet{Cas16} took great care to properly continuum-normalize the spectra used in the Cannon, arguing that standard continuum normalization techniques are signal-to-noise dependent, and small errors in continuum normalization can lead to large errors in label estimates. The flux-slope, however, should be less sensitive to such errors, and completely insensitive to constant offsets in the continuum.}

In order to perform the differentiation, we used a Savitsky-Golay (S-G) filter. We chose the S-G filter because it can efficiently smooth and differentiate the data simultaneously. Another desirable feature of the S-G filter is that, unlike most other smoothing filters, it can preserve the higher order moments of spectral lines. We used the S-G filter to calculate the slope and slope error at each pixel based on a window which could be arbitrarily wide. We ran the analysis using a 3-pixel window and a second degree S-G polynomial. The 3-pixel window corresponds to $\Delta\vsini \sim \unit[7]{\kms}$. \nt{This minimum S-G window width is the most physically sensible choice given the range of \vsinie spanned by the data. Repeating the analysis with larger windows (5-, 7- and 9-pixels) yielded inferior results when comparing our \vsinie estimates to those generated by ASPCAP.} The calculation of slope and slope error using the S-G filter is described in detail in Appendix \ref{sec:sgfilter}.

%Rotation, however, broadens and blends spectral lines, making the flux at each pixel more and more covariant. At some degree of rotation, treating pixels independently will not be sufficient.  
% Savgol window perhaps addresses the pixel-independence limitation?

The framework for our approach is as follows. While \citet{Ness15} formulate a model that describes the flux at each wavelength (Equation 1, in their paper):

\begin{eqnarray}
f_{n\lambda} &=&
g(\starlabelvec_n |  \set{\theta}_\lambda) + \mbox{noise}
\label{eq:specmodel}\quad 
\end{eqnarray}

we write an analogous model that instead describes the slope of the spectral flux at each wavelength:

\begin{eqnarray}
f^\prime_{n\lambda} &=&
g(\starlabelvec_n |  \set{\theta}_\lambda) + \mbox{noise}
\label{eq:specmodel_prime}\quad 
\end{eqnarray}

We can write a linear form without much loss of generality, similar to their Equation 2:

\begin{equation}
    f^\prime_{n\lambda} = \boldsymbol{\theta}^{\boldsymbol{T}}_\lambda \cdot \starlabelvec_n + \textrm{noise}
\end{equation}

where $\starlabelvec_n$ is the $k$-element label vector belonging to star $n$, and $\set\theta^T_\lambda$ is the \nt{$k$-element} vector of coefficients. The noise term is:

\begin{equation}
    \textrm{noise} = \xi (\sigma^{\prime\,2}_{n\lambda} + \scatter^2)
\end{equation}

where $\sigma^{\prime\,2}_{n\lambda}$ is the uncertainty of the slope for star $n$ at wavelength $\lambda$, $\scatter^2$ is the intrinsic scatter of the model at $\lambda$, and $\xi$ is a Gaussian random number. \nt{We proceed in two steps. In Section \ref{subsec:training_step}, we describe how we train the model at each wavelength by optimizing for the coefficients, $\set\theta_\lambda$, and the scatter term, $\scatter^2$. Then, in Section \ref{subsec:test_step}, we describe how we use the trained model to determine \vsinie from the flux-slope data alone.}

%% Training step
\subsection{Training Step}
\label{subsec:training_step}
Once we have determined a suitable form for our linear model (see Section \ref{subsec:model_selection}), we train the model parameters $\bm{\theta}_{\lambda}$ using the reference data from the California Kepler Survey. Like \citet{Ness15}, we build successively more complicated models in order to determine the necessary level of complexity.
The simplest model is one which is linear in the labels:

\begin{eqnarray}
\starlabelvec_n &\equiv& \left[1,
                           \frac{\starlabel_{n1} - \mean{\starlabel_1}}{2\sigma_{\starlabel 1}},
                           \frac{\starlabel_{n2} - \mean{\starlabel_2}}{2\sigma_{\starlabel 2}},
                           \cdots,
                           \frac{\starlabel_{nk} - \mean{\starlabel_k}}{2\sigma_{\starlabel k}}\right]
\label{eq:linear}\quad.
\end{eqnarray}

Here, the ``$1$'' is the intercept term, and subtracting the mean of each label `centers' the data. The division by $2$ standard deviations `standardizes' the data, so that labels with larger dynamic ranges do not dominate the regression. We denote standardized, centered labels with the hat notation, $\hat{\starlabel}_{nk} \equiv \frac{\starlabel_{nk} - \mean{\starlabelvec_k}}{2\sigma_{\starlabelvec k}}$. While the model is linear, the label vector need not be linear in the labels. In fact, it can be composed of any set of arbitrary functions of the labels. The model will nevertheless generate a flux-slope from a linear combination of these functions, weighted by the coefficients, $\set{\theta}_\lambda$. On the other hand, the \nt{problem} will become non-linear \nt{in the estimation step}, as the elements of the label vector will no longer be linearly independent in label space.

We know the $f^\prime_{n\lambda}$ and $\sigma^\prime_{n\lambda}$ from the apStar spectra, and the $\starlabelvec_n$ from the CKS data, so we are solving for $\set{\theta}^T_\lambda$ and $\scatter^2$. The log likelihood of the slope data in spectrum $n$ at pixel $\lambda$ is:

\begin{eqnarray}
\ln p(f^\prime_{n\lambda}\given\set{\theta}^T_\lambda, \starlabelvec_n, \scatter^2) &=&
 -\frac{1}{2}\,\frac{[f^\prime_{n\lambda} - \set{\theta}^T_\lambda \cdot \starlabelvec_n]^2}{\scatter^2 + \sigma^{\prime\,2}_{n\lambda}}
 -\frac{1}{2}\,\ln(\scatter^2 + \sigma_{n\lambda}^{\prime\,2})
\label{eq:like}\quad.
\end{eqnarray}
At fixed $\scatter^2$, the maximum-likelihood $\set{\theta}^T_\lambda$ can be solved for through straightforward linear algebra operations. For simplicity of implementation, we optimize for $\scatter^2$, and the resulting $\set{\theta}^T_\lambda$, using the IDL $\texttt{AMOEBA}$ code, which performs a downhill simplex optimization \citep{NM1965}.  

In practice, at each pixel $\lambda$, $\texttt{AMOEBA}$ iteratively solves the matrix equation,

\begin{equation}
    \label{eq:trainlinalg}
    \bm{M}\set{\theta}_\lambda = \bm{Y}_\lambda
\end{equation}
where $\bm{M}$ is the design matrix whose rows are the label vectors of each star, $\set{\theta}_\lambda$ is the set of regression coefficients at pixel $\lambda$, and $\bm{Y}_\lambda$ is the data vector, composed from each star's spectral slope at pixel $\lambda$. More explicitly, 

\begin{eqnarray}
\label{eq:trainmtx}
    \left[\begin{array}{llll}
        1&\ \hat{\starlabel}_{n=1,k=1}&\ \cdots\ &\ \hat{\starlabel}_{n=1,k=k}\\
        1&\ \hat{\starlabel}_{2,1}&\ \cdots\ &\ \hat{\starlabel}_{2,k}\\
        &\vdots&\ \ddots&\ \ \vdots\\
        1&\ \hat{\starlabel}_{N,1}&\ \cdots\ &\ \hat{\starlabel}_{N,k}
    \end{array}\right] &
    \left[\begin{array}{l}
        \theta_{0}\\
        \theta_{1}\\
        \,\vdots\\
        \theta_{k}
    \end{array}\right]_\lambda = &
    \left[\begin{array}{l}
        f^\prime_{n=1}\\
        f^\prime_{2}\\
        \ \vdots\\
        f^\prime_{N}
    \end{array}\right]_\lambda
\end{eqnarray}

With the assumptions of Gaussian likelihoods and no covariance between pixels (as in Equation  \ref{eq:like}), the solution to this equation is:
\begin{equation}
    \label{eq:theta_in_training}
    \set{\theta}_\lambda = \left[\bm{M}^T\bm{C}_\lambda^{-1}\bm{M}\right]^{-1} \bm{M}^T \bm{C}_\lambda^{-1} \bm{Y}_\lambda
\end{equation}
where $\bm{C}$ is the covariance matrix with diagonal entries:
\begin{equation}
    \textrm{diag}\left(\bm{C}_\lambda\right) = \left[ \sigma^{\prime\,2}_1 + s^2, \sigma^{\prime\,2}_2 + s^2, \cdots, \sigma^{\prime\,2}_N + s^2 \right]_\lambda
\end{equation}

Each $\texttt{AMOEBA}$ iteration takes a trial value of $\scatter^2$, which determines the covariance matrix $\bm{C}$. The coefficient vector $\set{\theta}_\lambda$ is then given by Equation \ref{eq:theta_in_training}. The optimization converges on the value of $\scatter^2$ (and the resulting $\set{\theta}_\lambda$) that maximizes the likelihood of the data, given the model. The scatter term represents the accuracy of the model at each wavelength. Wavelengths at which the data have large scatter about the model will be down-weighted by the additional error term. 
%In this way, all usable pixels in the spectra are incorporated into the optimization.

The above optimization is carried out at each pixel, except those in the gaps between the three \apogee~ chips. Pixels with little usable data (i.e., those masked by the APOGEE bitmasks in most spectra), pixels where the $\texttt{AMOEBA}$ optimization does not converge, \nt{and pixels where \texttt{AMOEBA} converges to $\scatter = 0$} are flagged and masked. We set their coefficient vectors to $\bm{0}$, which enforces that \nt{the model is insensitive to the input labels at those wavelengths}. \nt{Furthermore}, we set their scatter terms to a high value ($10^3$ times the scatter value of the worst good pixel). In this way, these bad pixels are automatically incorporated into the generative model, but do not contribute any information.

At the end of the training step, a coefficient vector and a scatter term has been derived for each pixel. \nt{This constitutes the generative model. In the estimation step, we determine the labels that generate the best-fit model to the observed spectra, given the trained model.}

%% Test Step
\subsection{\nt{Estimation} Step}
\label{subsec:test_step}
%These trained data are then used in the test step, where 
\nt{In the estimation step,} we seek to determine the label vector $\starlabelvec_n$ \nt{for stars without reference labels. Again, we maximize the log likelihood described in Eq. \ref{eq:like}, but} rather than \nt{summing the likelihood over the training spectra} at each pixel, we now \nt{sum the likelihood over all pixels for each unlabeled spectrum. That is, in contrast to the matrix representation in Eqns. \ref{eq:trainlinalg} and \ref{eq:trainmtx}, we have:}
\begin{equation}
    \bm{\Theta} \starlabelvec_n = \bm{Y}_n
\end{equation}
where $\bm{\Theta}$ is the design matrix with each row $i$ composed of the derived $\set{\theta}^T_{\lambda=i}$.  The matrices are of the form:

\begin{eqnarray}
    \left[\begin{array}{llll}
        \theta_{0, \lambda=1}   &\ \theta_{1, \lambda=1} & \cdots  &\ \theta_{k, \lambda=1}\\
        \theta_{0, 2}   &\ \theta_{1, 2} & \cdots  &\ \theta_{k, 2}\\
        \ \vdots&\ \ \vdots&\ddots&\ \ \vdots\\
        \theta_{0, \textrm{Npix}}   &\ \theta_{1, \textrm{Npix}} & \cdots &\ \theta_{k, \textrm{Npix}}
    \end{array}\right] &
    \left[\begin{array}{l}
        1\\
        \hat{\starlabel}_1\\
        \,\vdots\\
        \hat{\starlabel}_k
    \end{array}\right]_n = &
    \left[\begin{array}{l}
        f^\prime_{\lambda=1}\\
        f^\prime_{2}\\
        \ \vdots\\
        f^\prime_{\textrm{Npix}}
    \end{array}\right]_n
\end{eqnarray}
In the simplest case of a linear-in-labels model, the labels can again be determined by ordinary linear algebra. For more complicated models, such as the quadratic-in-labels model, this is not the case. 
 As a general solution, we again use \texttt{AMOEBA} to optimize the labels. Each iteration of \texttt{AMOEBA} takes a trial set of stellar label values and constructs the $\starlabelvec_n$, including non-linear terms, if necessary. The model is generated from the matrix multiplication of the label vector and the coefficient matrix from the training step. We calculate the likelihood of the data, given the model, and converge upon the set of labels that maximize the likelihood. We adopt the \vsinie from the best-fit label vector as our \vsinie estimate.
 
\subsection{Model Selection \nt{Using Cross-Validation}}
\label{subsec:model_selection}

\cite{Ness15} found that for their implementation of The Cannon, the simplest sufficient model was quadratic in three labels (\teff, \feh, and \logg), which was a label vector of the form:

\begin{eqnarray}
\starlabelvec_n &=& \begin{array}{l}[1,
                          \hat{\starlabel}_{n1}^2,
                          \hat{\starlabel}_{n2}^2,
                          \hat{\starlabel}_{n3}^2,\\
                          \hat{\starlabel}_{n1}\hat{\starlabel}_{n2},
                          \hat{\starlabel}_{n2}\hat{\starlabel}_{n3},
                          \hat{\starlabel}_{n3}\hat{\starlabel}_{n1},\\
                          \hat{\starlabel}_{n1},
                          \hat{\starlabel}_{n2},
                          \hat{\starlabel}_{n3}]\quad .
\end{array}
\label{eq:quadinlabels}
\end{eqnarray}

%%Aesthetically, we felt it would be cleaner and more compact to incorporate \vsinie directly into our model.

\nnt{In order to incorporate rotation into the model, we added \vsinie as a label in the label vector. We note that it should also be possible to exclude \vsinie from the label vector, and instead to generate models using only the three fundamental parameters, explicitly broadening the spectra in a separate step through convolution with a broadening kernel. That method would also be efficient, but we decided to include \vsinie as a label for a number reasons.  We envisioned our approach as a method to take a set of ``\vsinie standards'' (the CKS stars), and propagate those measurements to other data sets. Because there is no ``standard'' way to measure \vsini, different techniques can yield different results (e.g., due to different assumed values of the wavelength-dependent limb darkening coefficient, which has a small, but potentially non-negligible effect on the shape of the broadening kernel). Including \vsinie in the label vector allows us to avoid introducing any additional assumptions about the underlying physics, while empirically estimating \vsinie for potentially heterogeneous data sets in a self-consistent way.}

With this approach in mind, we tested the model's ability to recover \vsinie using a $10$-fold cross-validation \citep{Ive14} on the two training samples described in Section \ref{sec:data}. We tested many combinations of those $3$ fundamental parameters and \vsini, along with a number of functions of the labels. Ultimately, we achieved the best results with a label vector that is quadratic in the three labels intrinsic to the star (identical to the Cannon's label vector in Equation \ref{eq:quadinlabels}) and independently quadratic in the extrinsic label \vsini. In other words, denoting \vsini~as label four, our label vector is:

\begin{eqnarray}
%\starlabelvec_n &\equiv& \begin{array}{l}[1,\hat{\starlabel}_{n4}, \hat{\starlabel}_{n4}^2,\\
%                          \hat{\starlabel}_{n1}^2,
%                          \hat{\starlabel}_{n2}^2,
%                          \hat{\starlabel}_{n3}^2,\\
%                          \hat{\starlabel}_{n1}\hat{\starlabel}_{n2},
%                          \hat{\starlabel}_{n2}\hat{\starlabel}_{n3},
%                          \hat{\starlabel}_{n3}\hat{\starlabel}_{n1},\\
%                          \hat{\starlabel}_{n1},
%                          \hat{\starlabel}_{n2},
%                          \hat{\starlabel}_{n3}]\quad .
\starlabelvec_n &\equiv& \begin{array}{l}[\starlabelvec_{\textrm{Eq.} \ref{eq:quadinlabels}},\hat{\starlabel}_{n4}, \hat{\starlabel}_{n4}^2]\quad .
\end{array}
\label{eq:finalmodel}
\end{eqnarray}

\begin{figure}[ht] %Partial Derivatives
    \includegraphics[width=1\textwidth]{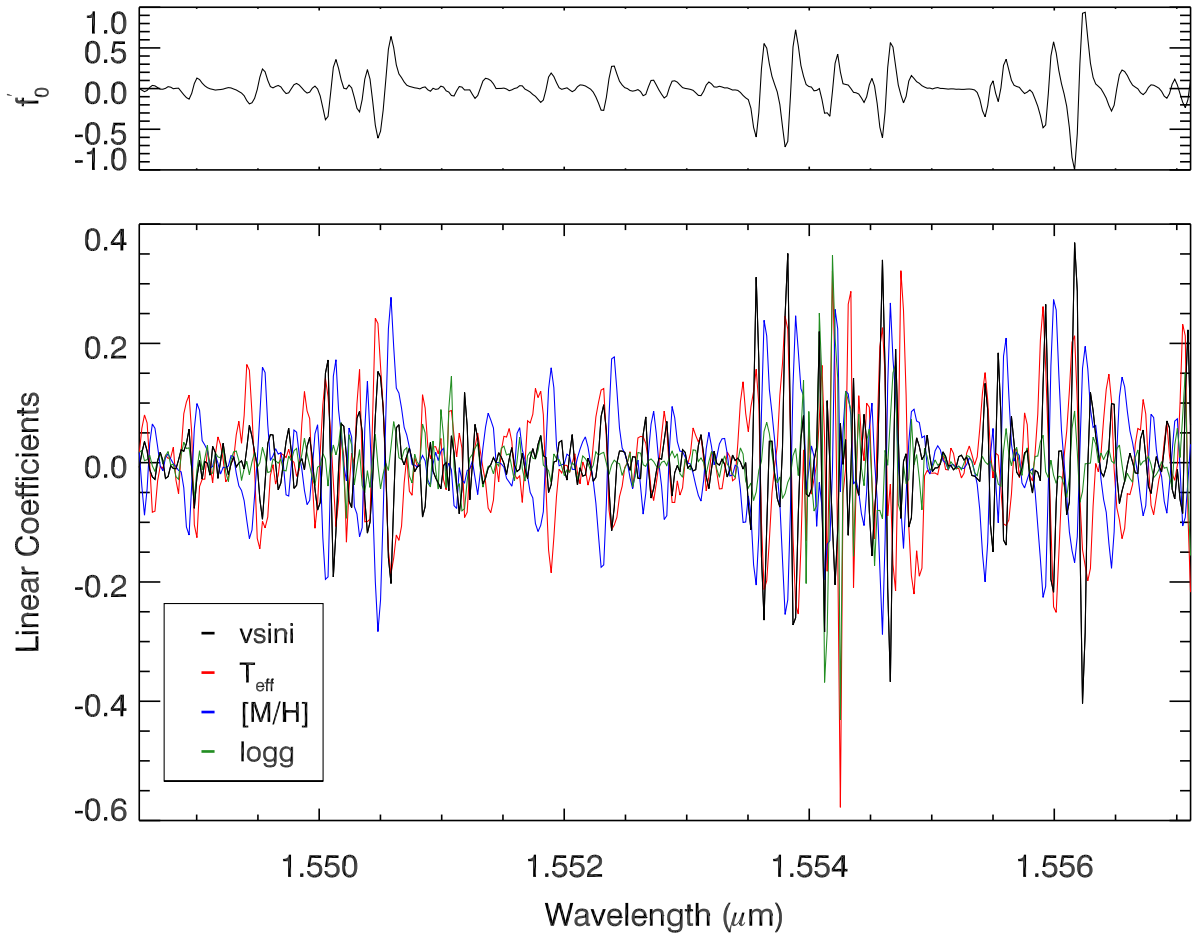}
    \caption{A portion of the model and its first-order derivatives with respect to the labels. The upper panel is the flux-slope model generated from the median labels. The lower panel shows the linear coefficients for each of the labels, color-coded by label.}
    \label{fig:derivs}
\end{figure}

\begin{figure}[ht] %Vsini coeff vs. empirical derivative
    \includegraphics[width=1\textwidth]{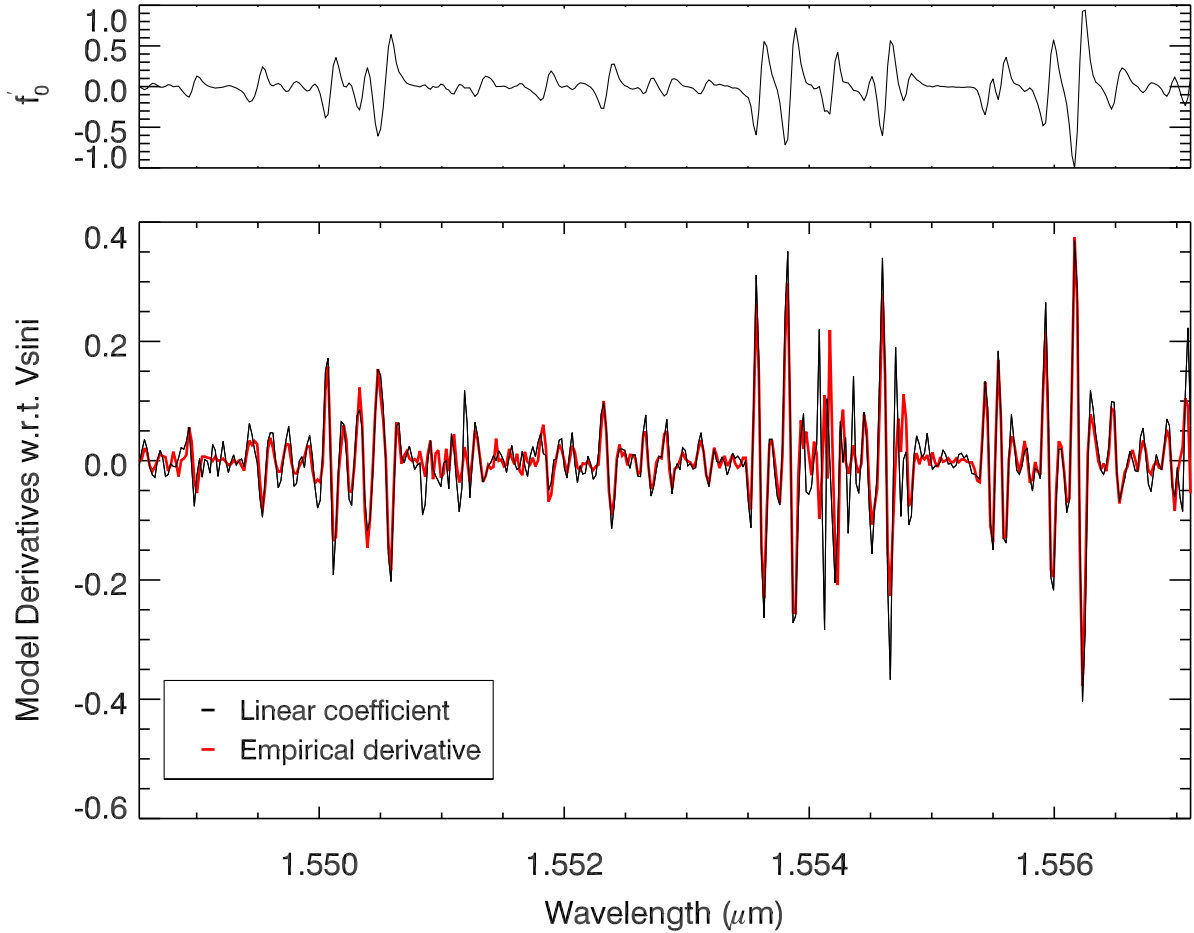}
    \caption{A comparison between the first-order \vsinie coefficient and the empirical first-order derivative of the median model with respect to \vsini. The upper panel is the flux-slope model generated from the median labels. In the lower panel, the model's first-order \vsinie coefficient is plotted in black, and the partial derivative of the model with respect to \vsinie is overplotted in red. We determined the derivative empirically by broadening the median model spectrum by a small amount ($\Delta\vsini = \unit[0.2]{\kms}$), taking the difference between the broadened and unbroadened spectra, and dividing by the broadening amplitude. The agreement between the model \vsinie coefficient and the actual derivative with respect to \vsinie is an indication that the model is performing as expected.}
    \label{fig:vsinideriv}
\end{figure}

\nnt{We explore some aspects of the trained model in Figures \ref{fig:derivs} and \ref{fig:vsinideriv}. The upper panels of the figures show a portion of the median model flux-slope spectrum (i.e., the spectrum generated from a label vector with each label set to its median value.) In the lower panel of Figure \ref{fig:derivs}, we show the linear coefficients of the trained model, color-coded by label. The magnitudes of the coefficients give an indication of the relative dependence of the model on each label at a given wavelength. Unsurprisingly, the \vsinie coefficient is most pronounced at wavelengths corresponding to the sharpest spectral lines.}

\nnt{In the lower panel of Figure \ref{fig:vsinideriv}, we compare the linear \vsinie coefficient of the model to the empirical first-order derivative of the median model with respect to \vsini. The \vsinie coefficient is plotted in black, and the partial derivative of the model with respect to \vsinie is overplotted in red. We determined the derivative empirically by broadening the median model spectrum by a small amount ($\Delta\vsini = \unit[0.2]{\kms}$), taking the difference between the broadened and unbroadened spectra, and dividing by the broadening amplitude. The agreement between the model \vsinie coefficient and the actual derivative with respect to \vsini, which should be approximately equal, is an indication that the model is performing as expected.}
 
\section{Results}

\subsection{Training Sample \nt{Cross-Validation Results}}
\begin{figure}[ht] %Vsini Result from cross-validation
    \includegraphics[width=1\textwidth]{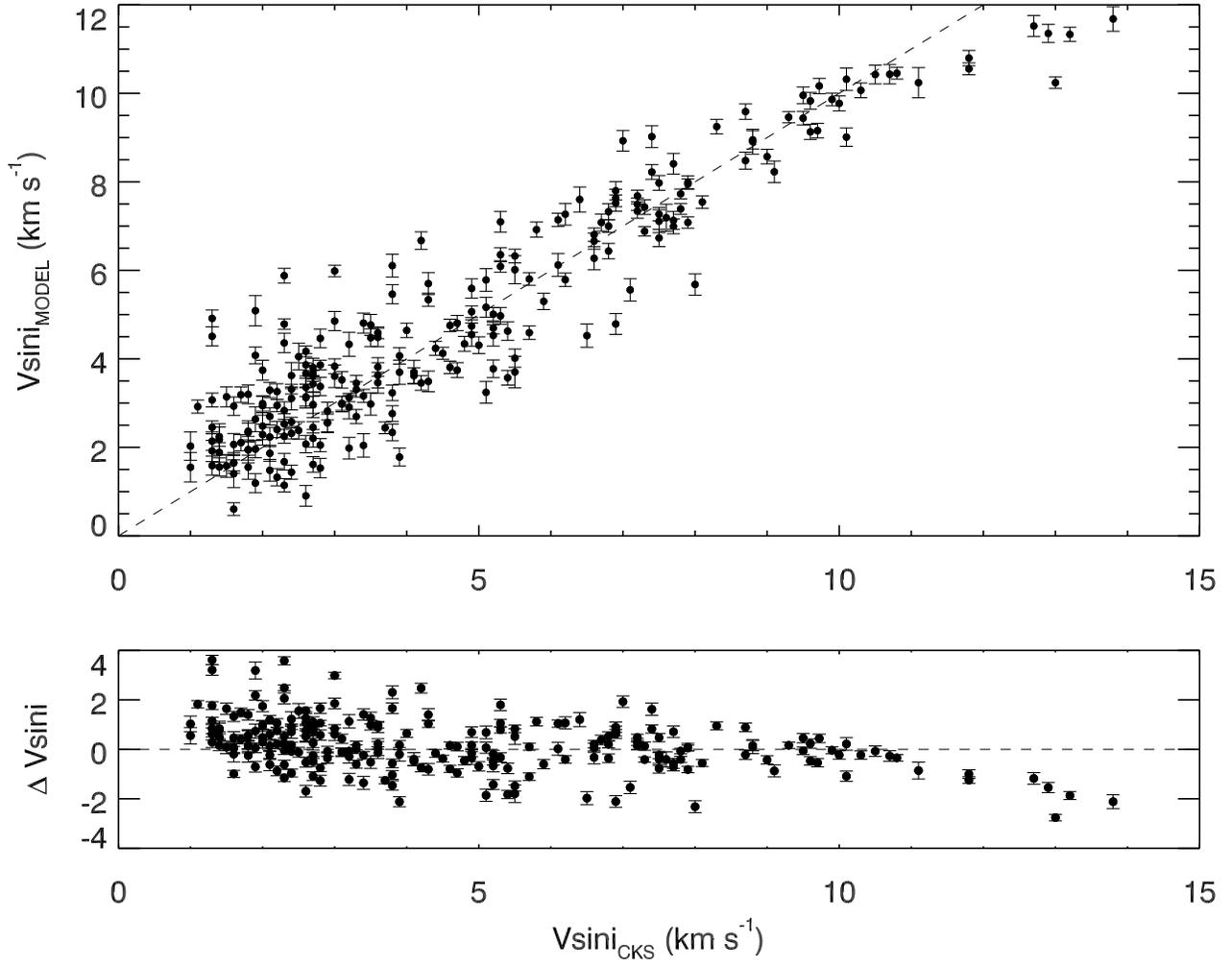}
    \caption{Results of $10$-fold cross-validation on the subset of the training sample that contains only the stars which were measured to be detectably rotating by CKS. The model is quadratic in the three fundamental labels, and independently quadratic in \vsini. Due to the sparsity of high \vsinie training data, the model is noticeably biased above $\vsini \sim \unit[9]{\kms}$.}
    \label{fig:vsini_detonly}
\end{figure}

\begin{figure}[ht] %Vsini Result from cross-validation with fakes
    \includegraphics[width=1\textwidth]{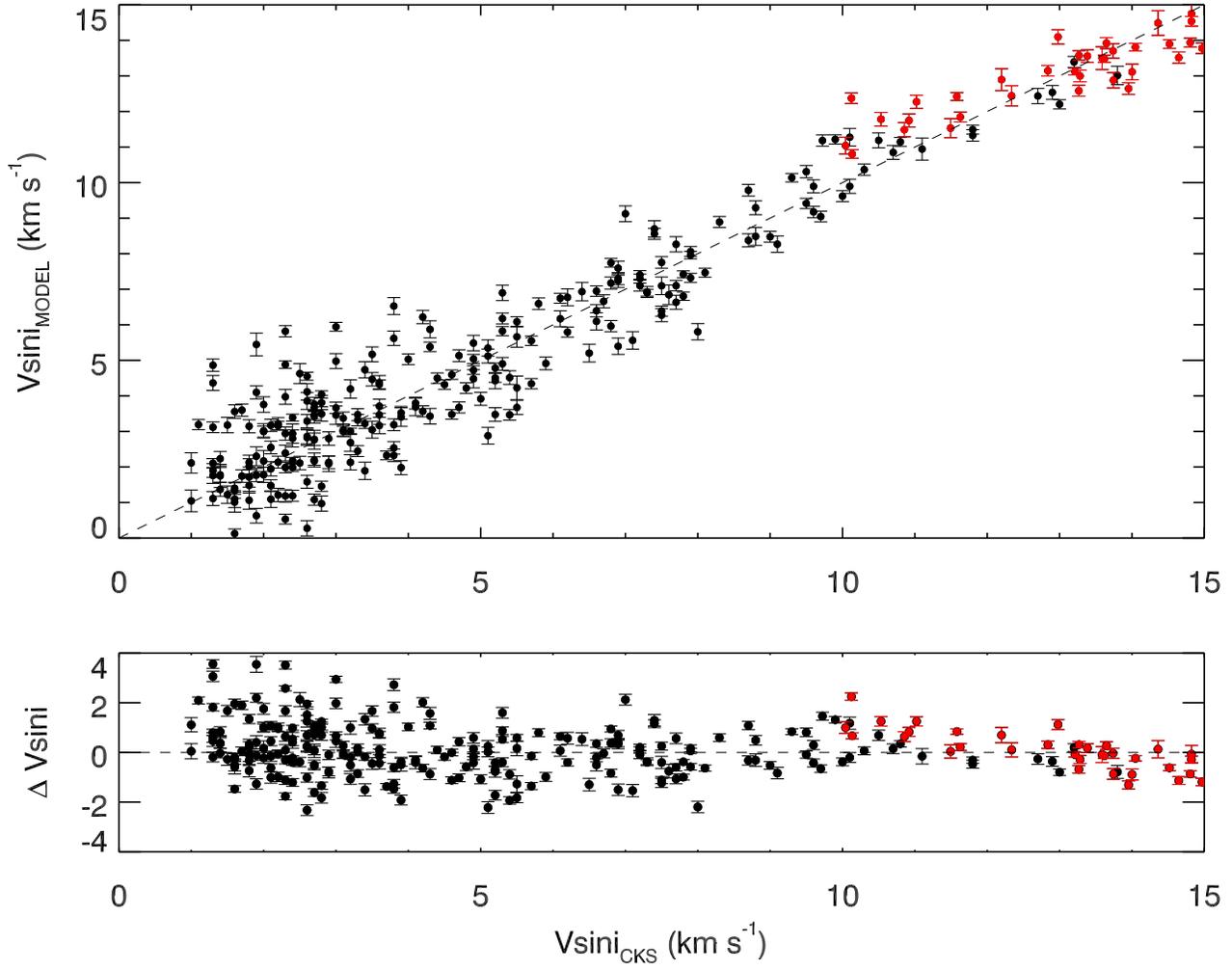}
    \caption{Results of $10$-fold cross-validation on the full training sample, including artificially broadened spectra of non-rotators. The broadened spectra are plotted in red. The addition of data with \vsinie between \unit[10]{\kms} and \unit[15]{\kms} improves the model's performance at high \vsini. }
    \label{fig:vsini_wdets}
\end{figure}

The \vsinie results of our cross-validation on the training samples are shown in Figures \ref{fig:vsini_detonly} and \ref{fig:vsini_wdets}. \nnt{In each of the $10$ folds of the cross-validation, we trained the model on a different $90\%$ of the training data, and used the model to estimate \vsinie for the remaining $10\%$. The \vsinie estimates calculated for each star, when left out of the training set, are plotted against the CKS reference measurements.} The error bars in both figures are the formal uncertainties derived from the parameter covariance matrix in the test step, and range from $\unit[0.1]{\kms}$ to $\unit[1]{\kms}$. These should be added in quadrature to the CKS \vsinie uncertainty of $\unit[1]{\kms}$. Figure \ref{fig:vsini_detonly} shows the results for the subset of the training sample without artificially broadened non-detections. Because the data is sparse at $\vsini \gtrsim \unit[9]{\kms}$, the high \vsinie results are biased. Figure \ref{fig:vsini_wdets} shows the results for the full training set, including the artificially broadened spectra. Adding the synthetic fast rotators (red points) dramatically improved the model's ability to recover large \vsini.   

\begin{figure}[ht] %Vsini cross-validation results binned
    \includegraphics[width=1\textwidth]{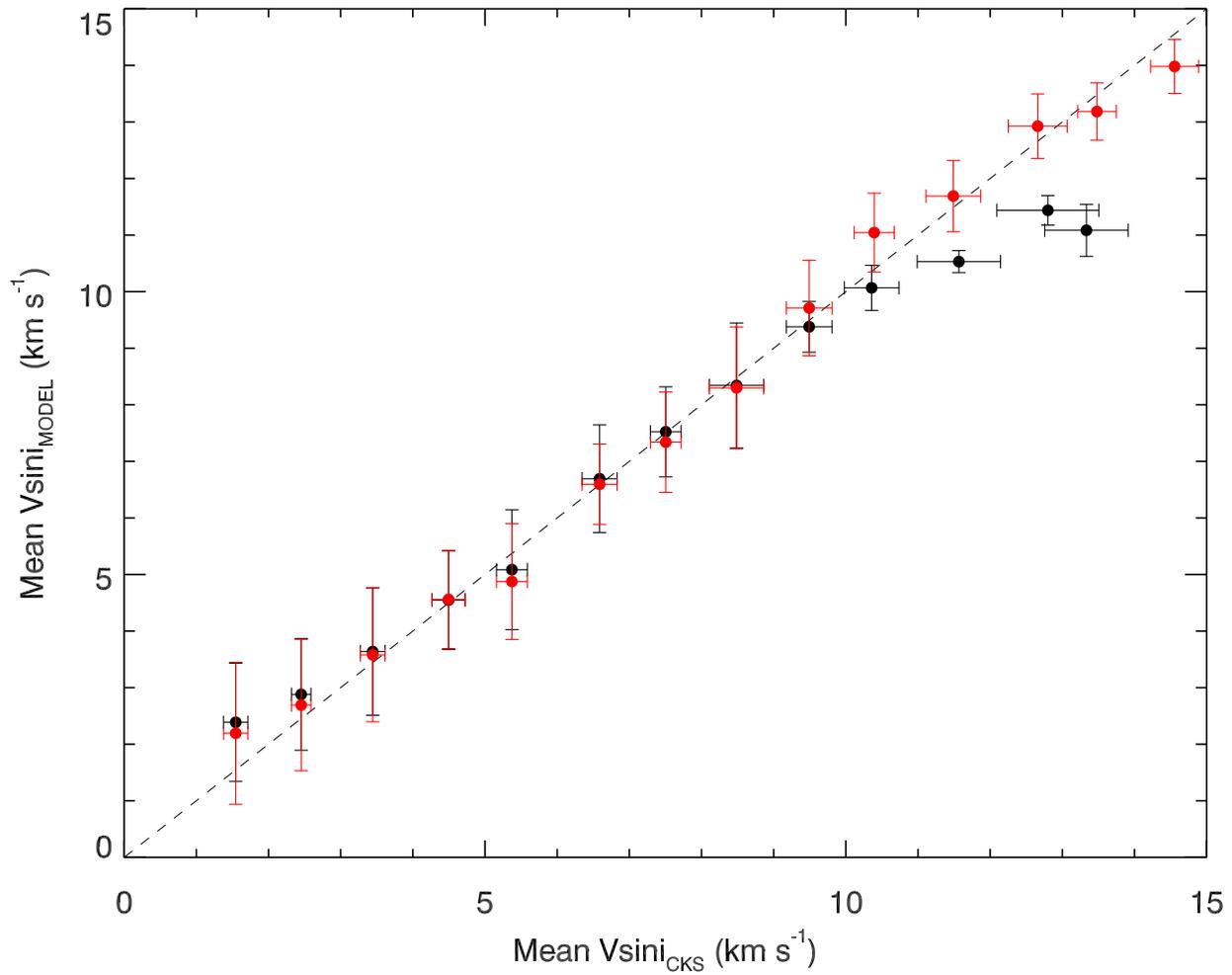}
    \caption{Results from cross-validation of the training sample, binned in units of $\unit[1]{\kms}$. The binned data from the training sample subset are plotted in black, and the data from the full training sample, including the artificially broadened spectra of non-rotators, are plotted in red. The horizontal error bars show the uncertainty in the CKS mean \vsini. The vertical error bars show the standard deviation of the residuals in each bin.}
    \label{fig:binned_detonly}
\end{figure}

This is also reflected in Figure \ref{fig:binned_detonly}, which summarizes the precision and accuracy of \vsinie estimated by our approach, as a function of \vsini. The data are binned by CKS \vsinie in $\unit[1]{\kms}$ increments, and the mean values of the model \vsinie are plotted versus the mean CKS \vsini. The horizontal error bars represent the uncertainty in the mean of the CKS \vsini, and the vertical error bars represent the scatter in the model \vsinie (standard deviation of the residuals). The data from the detections-only sub-sample are plotted in black, and the data from the full training sample are plotted in red. The scatter in the recovered \vsinie is approximately $\unit[1]{\kms}$ down to $\vsini \sim \unit[2]{\kms}$.

\begin{figure}[ht] %Test Results
    \includegraphics[width=1\textwidth]{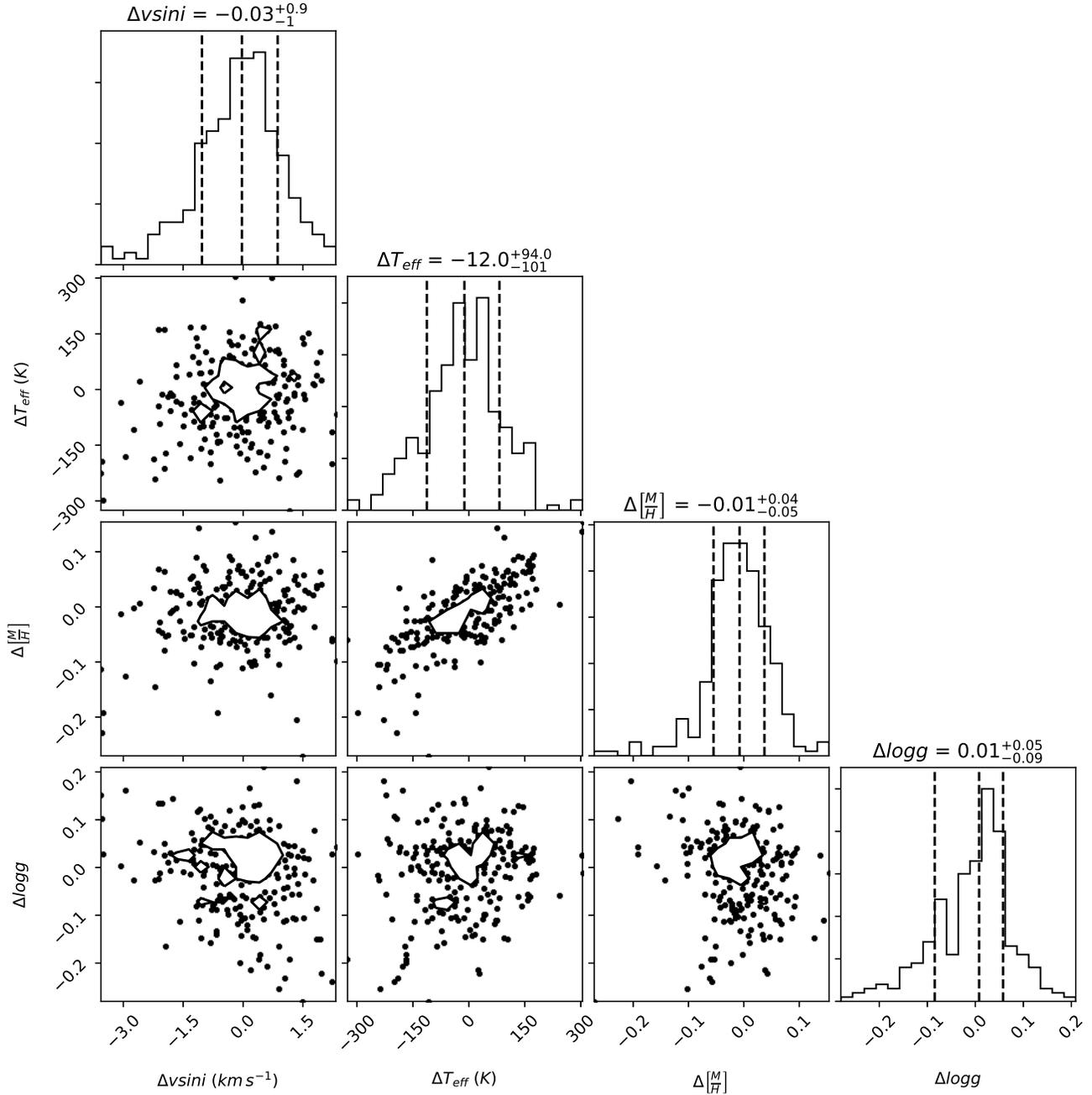}
    \caption{ Residuals from the 10-fold cross validation of the training data, plotted in various projections of the labels. The scatter plots show that the residuals are typically not correlated between labels. There does, however, appear to be some correlation in the \feh-\teffe plane. The contours denote $1$-$\sigma$ confidence levels. The residuals in each label are shown in the histograms. Overall, the residuals are only slightly biased, have symmetrical distributions, and uncertainty comparable to that of the input labels.  }
    \label{fig:crossval_results}
\end{figure}

\nt{The overall results of the 10-fold cross-validation of the training sample are summarized in Figure \ref{fig:crossval_results}. The residuals are plotted in various projections of label space. Other than in the \met-\teffe space, the residuals appear largely uncorrelated between labels. The distribution of residuals in each individual label are illustrated in the histograms. The distributions are apparently non-Gaussian, but still symmetric and unimodal. They exhibit only a small amount of bias, and have uncertainties similar to those of the CKS reference data.}

\begin{figure}[ht] %Data with model
    \plottwo{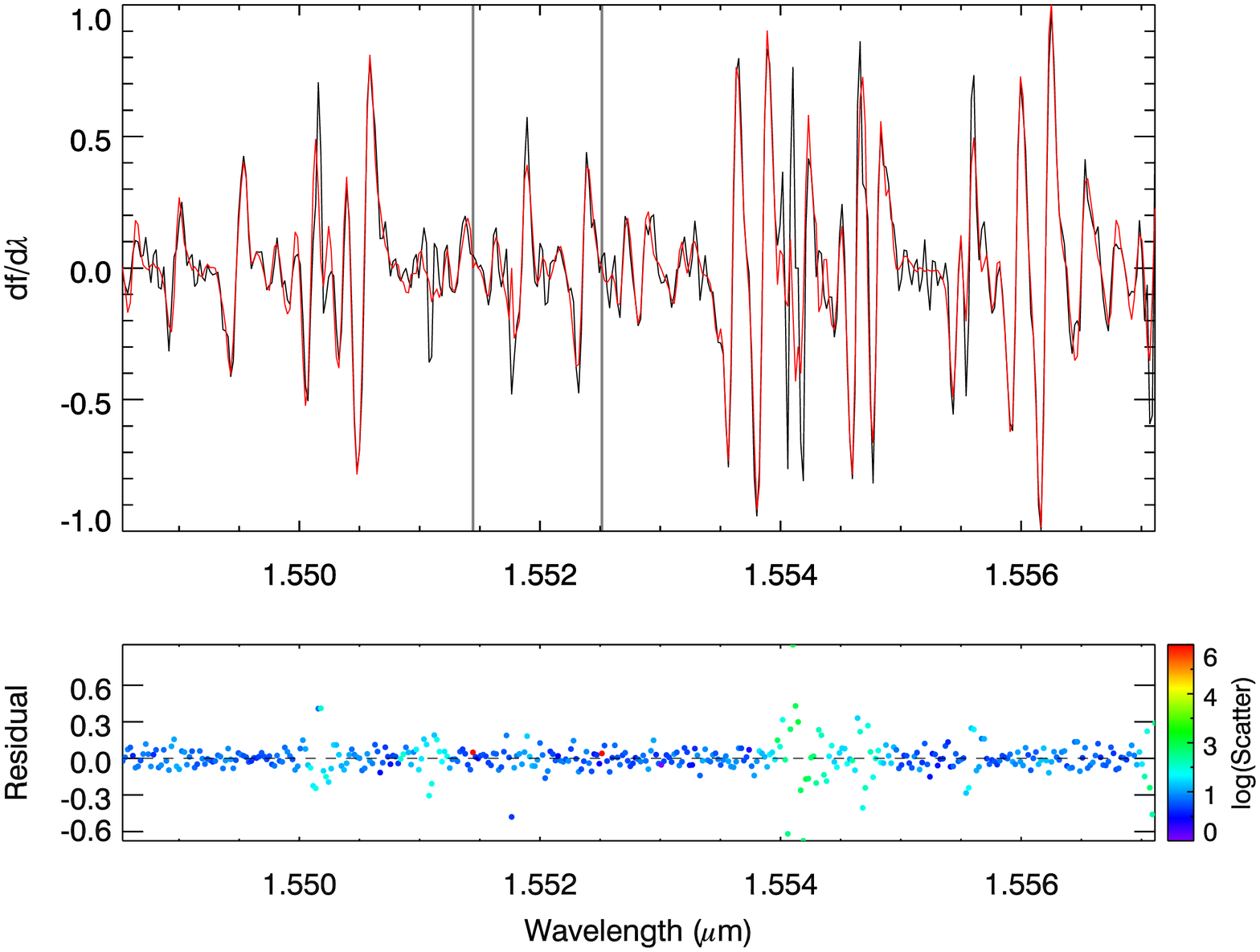}{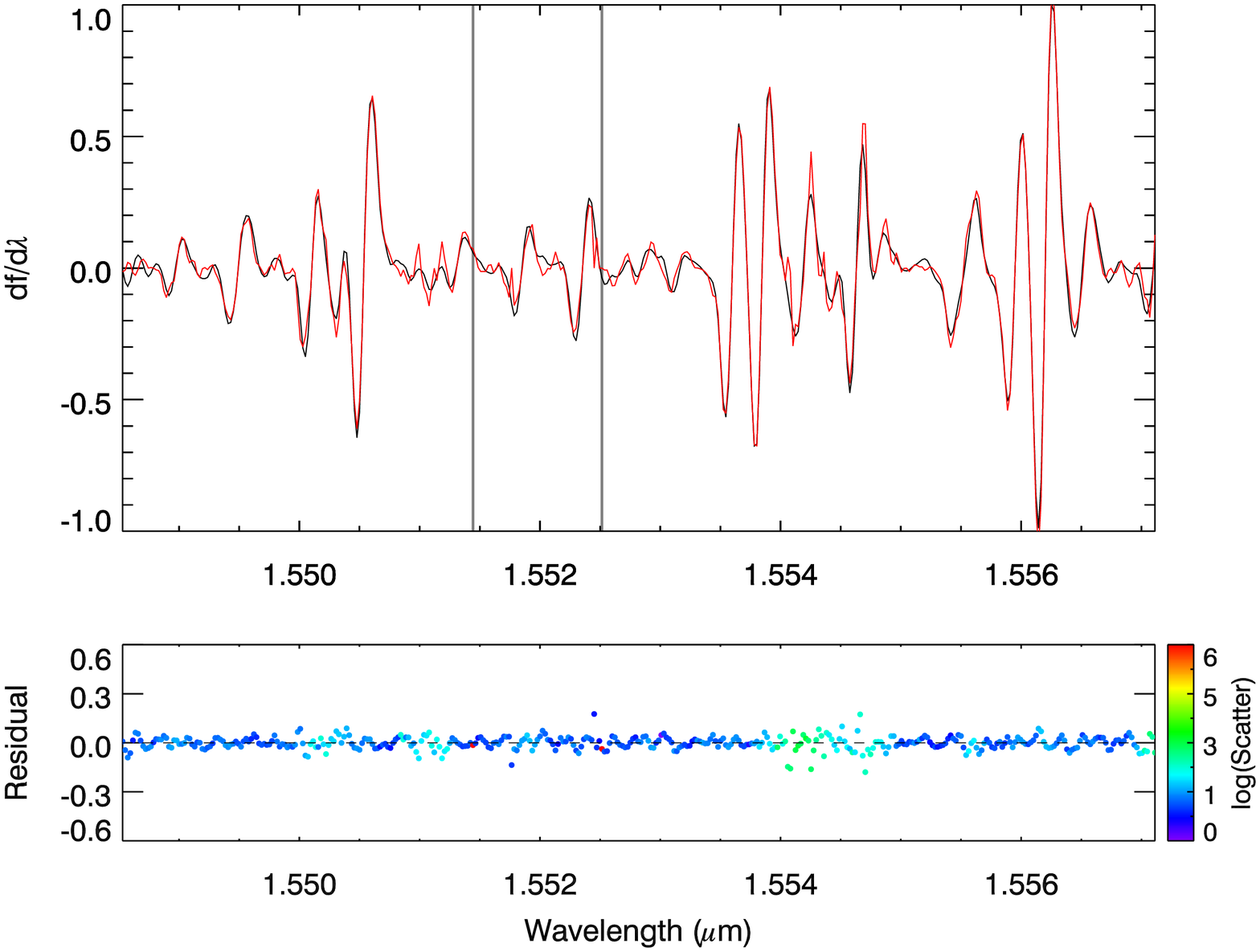}
    \caption{A portion of the spectral slope data, along with the generated model, for two of our training spectra. The spectral slope is arbitrarily scaled to range from $-1$ to $1$, and plotted in black in the upper panels. The model is generated by the trained parameters with the spectra in question left out from the training set. It is overplotted in red, with masked pixels highlighted by the vertical gray bands. The residuals are shown in the bottom panel, and color-coded by the scatter term associated with each wavelength. }
    \label{fig:data_model}
\end{figure}

\nt{We show two examples of our spectral flux-slope data in comparison with the generated models in Figure \ref{fig:data_model}. The spectral slope is displayed in the upper panel in black, arbitrarily scaled to range from $-1$ to $1$. The model spectra generated from the trained parameters (with these spectra held out from the training set), and from which we infer \vsini, is overplotted in red. The bottom panel shows the residuals, color-coded by the scatter term associated with each wavelength. Masked pixels are highlighted by vertical gray bands. The models generally fit the data well. Areas where the residuals are large have large scatter terms, which automatically down-weight those wavelengths in the likelihood calculation.} One star, KOI 3052 (not shown), appeared to be an obvious outlier. Details of what we observed and how we treated this star are described in Appendix \ref{sec:outlier}.

\subsection{\nt{Survey Sample Results}}
\begin{figure}[ht] %Vsini Results with Survey Data
    \includegraphics[width=1\textwidth]{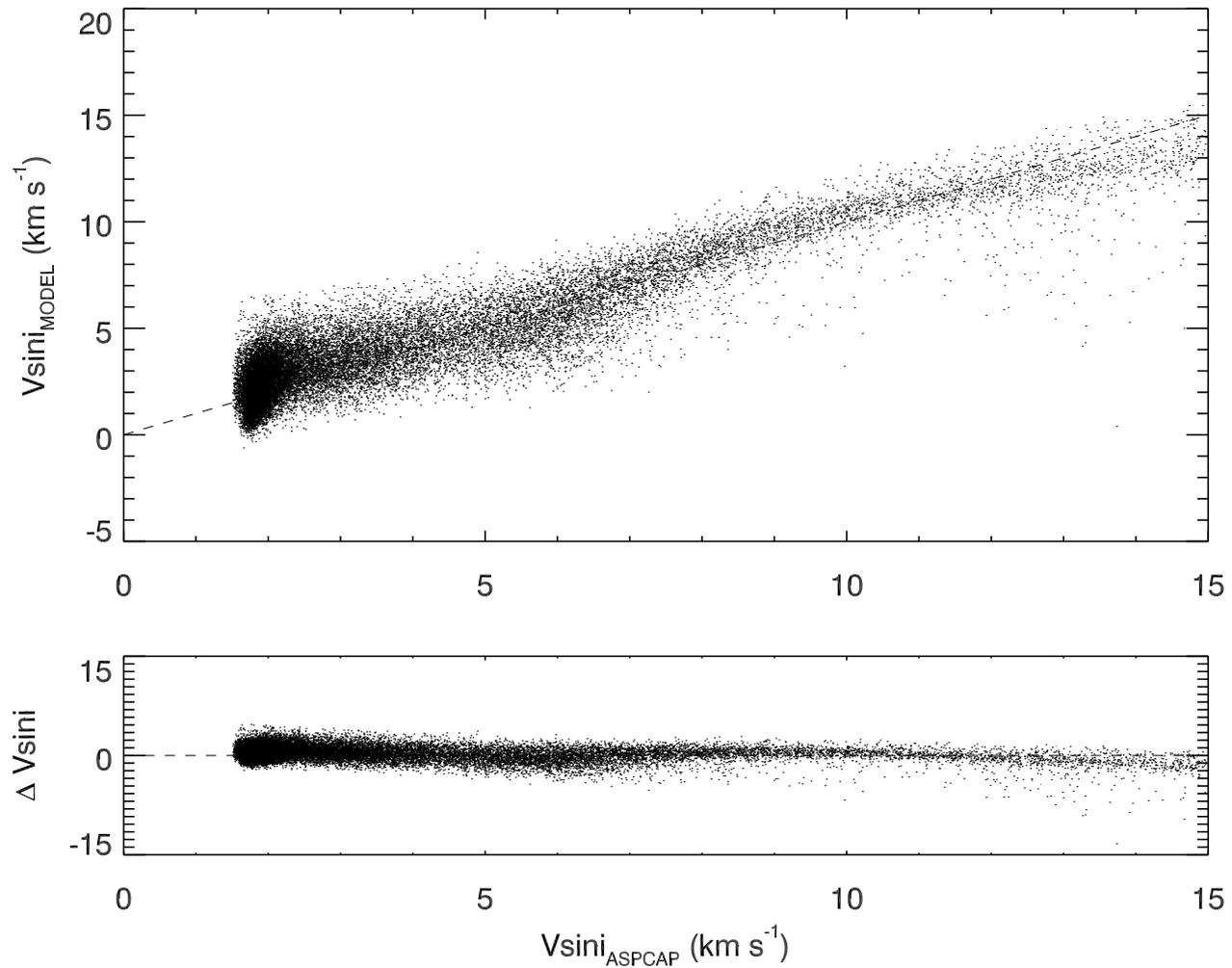}
    \caption{Results of the full analysis on $27,000$ APOGEE \nt{survey} spectra, using the full training sample of $270$ APOGEE/CKS stars. The APOGEE spectra and the CKS labels of the training stars are used to train the model. The trained model is then used to estimate \vsinie for the $27,000$ stars in the \nt{survey} sample. The model \vsinie estimates are plotted versus the ASPCAP \vsini. The RMS of the residuals is $\unit[1.2]{\kms}$}.
    \label{fig:boot}
\end{figure}

\begin{figure}[ht] %Chi2
    \plottwo{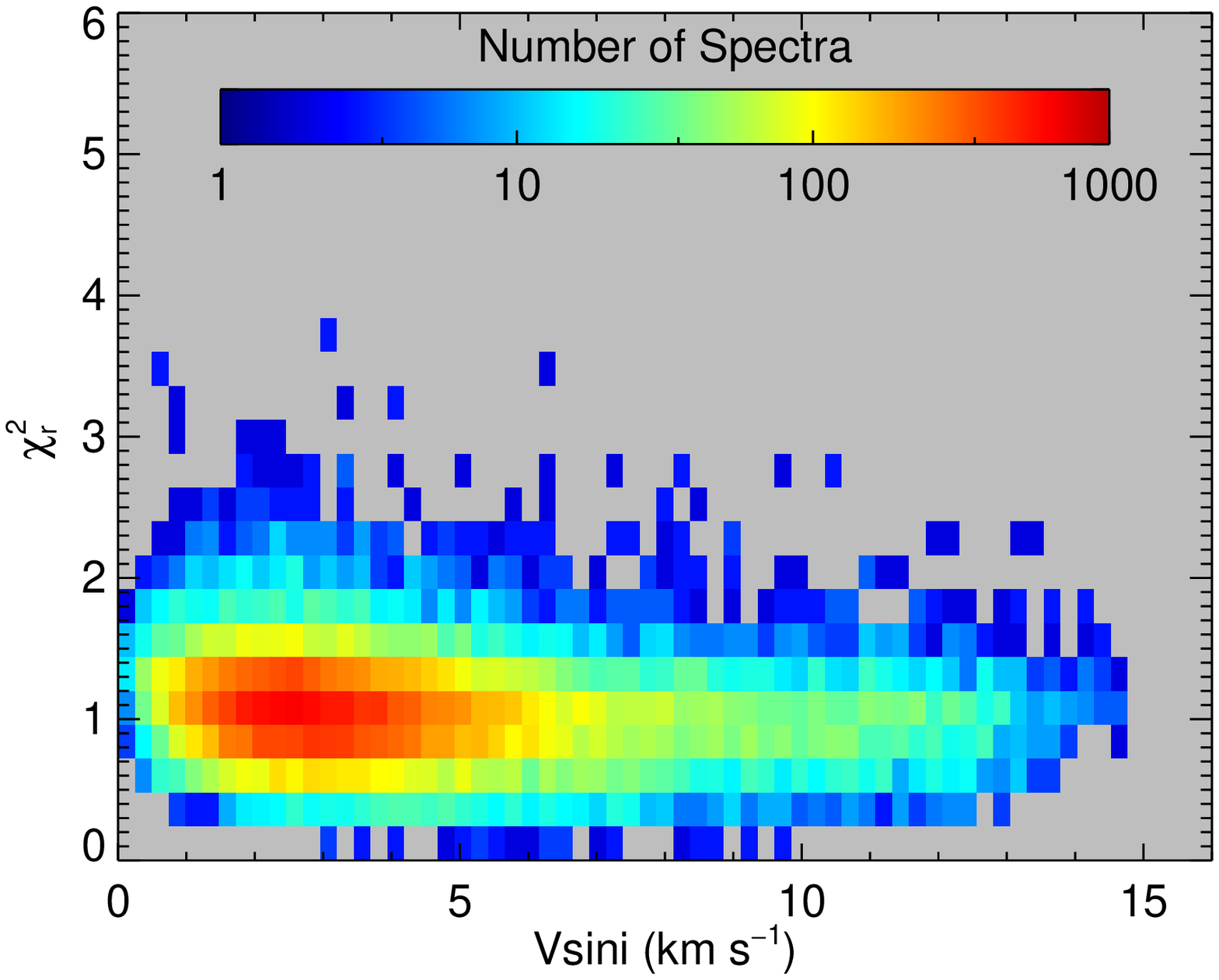}{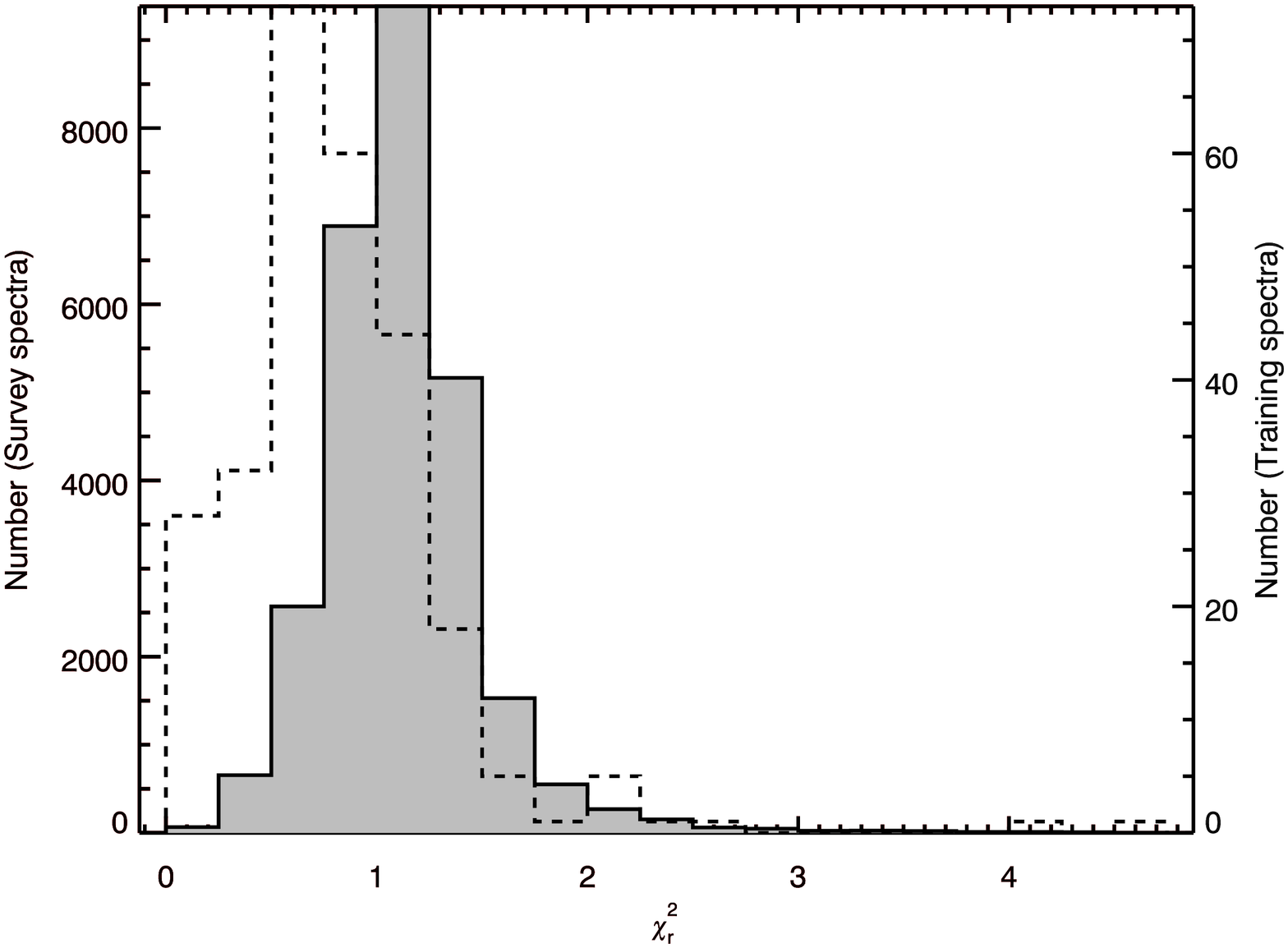}
    \caption{ The reduced $\chisq$ of the generated models. \textbf{Left:} The reduced $\chisq$ of the models from the APOGEE survey sample of $27,000$ stars, as a function of \vsini. The $\chisq$ distribution does not display any \vsini-dependence. \textbf{Right:} The overall reduced $\chisq$ distribution of both the training data models (generated in the 10-fold cross validation), and the full survey sample models. The dashed line indicates the training data distribution, and the gray-filled histogram indicates the survey sample distribution. }
    \label{fig:chi2}
\end{figure}

Once the model determination was complete, we carried out the full analysis. We trained the model parameters ($\set{\Theta}, \bm{s}^2$) using the full training sample, and then empirically determined \vsinie for the $27,000$ stars in the \nt{survey} sample. The \vsinie results are shown in Figure \ref{fig:boot}, plotted against the \vsinie estimated by ASPCAP. Although there are clear outliers, and there is structure in the residuals, the model \vsinie are generally consistent with the ASPCAP \vsini. The RMS of the residuals is $\unit[1.2]{\kms}$. \nt{In Figure \ref{fig:chi2}, we show that the reduced $\chisq$ exhibits no systematic dependence on \vsini. The right panel of Figure \ref{fig:chi2} shows the overall distribution of reduced $\chisq$ for the full survey sample. The dashed histogram shows the reduced $\chisq$ distribution for the training sample, obtained from the cross-validation step. In short, our method can reliably estimate \vsinie in the parameter space of the training sample at an uncertainty only slightly greater than that of the CKS reference \vsini. The reduced $\chisq$ of the models suggest that the flux-slope based generative model is good description of the APOGEE observations.} 

\section{Discussion and Conclusion}
% It's a new technique
This data-driven approach to \vsinie estimation is a powerful tool for the study of stellar angular momentum. Especially in this era of large scale spectroscopic surveys, our technique provides a \nt{(largely)} model-independent, computationally inexpensive means of measuring \vsinie for large numbers of stars at once. Using a UNIX workstation with four $3.4$ GHz CPUs, estimating \vsinie for the $27,000$ stars in the test sample took only $\unit[3.3]{hours}$ ($\sim \unit[0.44]{seconds}$ per spectrum), and roughly half of that time was spent on calculating the slope and slope error with our custom S-G filter, which \nt{could be optimized to perform faster}. This technique is orders of magnitude faster than fitting suites of theoretical templates, for example.
% Calibrate surveys and techniques (stated for CANNON)?
Potentially, this method could also be used to cross-calibrate data sets from different surveys, which may have conflicting measurements due to the use of different techniques or stellar models. 

% LIMITATION: No label errors 
We have already discussed some of the limitations of the framework we have chosen, such as the independent treatment of pixels which are, in reality, correlated. Our use of the spectral line slope, rather than spectral flux, is a shortcut that allowed us to capture some of the relationship between the flux at each pixel within this framework. The slope at each pixel incorporates flux information from a $3$-pixel window, which is roughly the width of a $\sim \unit[7]{\kms}$ broadening kernel at the APOGEE resolution and sampling. The model described here may not be sufficient at very high \vsini, where the broadening kernel is much larger than $3$-pixels. 
%\nt{Indeed, the fixed-width S-G window is a limitation }

\nt{Another possible limitation may be the use of polynomials as a basis for our model.} The broadening and blending of spectral lines results in a complicated relationship between \vsinie and line slope, which can vary widely between pixels. \nt{Although we found no functions of the labels that performed better than the quadratic model that we adopted, there is periodic structure in the residuals that is likely due to the polynomial form of the model.} Furthermore, it is not clear that a quadratic model would work well for data extending to higher \vsini. One possible improvement would be to use a Gaussian Process regression, which would obviate the need to specify a functional form for the model. The trade-off is that Gaussian Process regression is more computationally expensive (naively $\bigo(N^3)$). 

Another important feature to consider is that the training step does not incorporate label errors. \nt{This could adversely affect the accuracy of the model,} particularly at low \vsini, (e.g., $\vsini_{\textrm{CKS}} = 1 \pm \unit[1]{\kms}$), \nt{where} the label errors are \nt{a} non-negligible {fraction of the reference value}. A more sophisticated approach could treat the labels as probability distributions and propagate those uncertainties into posterior distributions through the use of Markov Chain Monte Carlo techniques. Again, such improvements would come at the cost of decreased speed.
% Limitation - training data needs to span label space

Finally, there are some limitations inherent to the data-driven aspect of this approach, regardless of the model complexity. One is the necessity of high-fidelity training data spanning label space. The amount of training data needed increases exponentially with the number of labels. Furthermore, the local density of training data within the spanned parameter space can limit the model's accuracy. In the case of \vsini, we have shown that artificially-broadened non-detections can be used to supplement the bona fide rotators at \vsinie where training data is sparse. 

In this work, we demonstrated a novel data-driven method for estimating \vsinie from stellar spectra. We used the first derivative of APOGEE spectra, along with high-fidelity parameter estimates from the California Kepler Survey, to train a generative model from which we estimated \vsinie for $27,000$ APOGEE F-G-K dwarfs. In the range $0 \le \vsini \le \unit[15]{\kms}$, the model produced \vsinie estimates that agreed with ASPCAP measurements to within $\unit[1.2]{\kms}$, in a fraction of the time required by standard \vsinie measurement techniques. 

%Another is the question of whether such a data-driven technique can be said to `measure' anything. It would be more precise to say that we \textit{propagated} the CKS labels to an unlabeled data set. The upshot of this subtlety is that the results of such a technique are necessarily coupled to the reference labels, and subject to the same systematic errors as the reference data.

\acknowledgements
We would like to thank our anonymous referee for comments that helped to significantly improve this manuscript. 

This work was supported in part by the Ella N. Pawling Endowment. This research has made use of NASA's Astrophysics Data System and the SIMBAD database, operated at CDS, Strasbourg, France \citep{Wen00}. 

Funding for the Sloan Digital Sky Survey IV has been provided by the Alfred P. Sloan Foundation, the U.S. Department of Energy Office of Science, and the Participating Institutions. SDSS acknowledges support and resources from the Center for High-Performance Computing at the University of Utah. The SDSS web site is www.sdss.org.

SDSS is managed by the Astrophysical Research Consortium for the Participating Institutions of the SDSS Collaboration including the Brazilian Participation Group, the Carnegie Institution for Science, Carnegie Mellon University, the Chilean Participation Group, the French Participation Group, Harvard-Smithsonian Center for Astrophysics, Instituto de Astrof\'isica de Canarias, The Johns Hopkins University, Kavli Institute for the Physics and Mathematics of the Universe (IPMU) / University of Tokyo, Lawrence Berkeley National Laboratory, Leibniz Institut f\"ur Astrophysik Potsdam (AIP), Max-Planck-Institut f\"ur Astronomie (MPIA Heidelberg), Max-Planck-Institut f\"ur Astrophysik (MPA Garching), Max-Planck-Institut f\"ur Extraterrestrische Physik (MPE), National Astronomical Observatories of China, New Mexico State University, New York University, University of Notre Dame, Observat\'orio Nacional / MCTI, The Ohio State University, Pennsylvania State University, Shanghai Astronomical Observatory, United Kingdom Participation Group, Universidad Nacional Aut\'onoma de M\'exico, University of Arizona, University of Colorado Boulder, University of Oxford, University of Portsmouth, University of Utah, University of Virginia, University of Washington, University of Wisconsin, Vanderbilt University, and Yale University.

\appendix
\section{Calculating Slope with the Savitsky-Golay Filter}
\label{sec:sgfilter}

 The S-G filter effectively performs a least-squares polynomial fit of order $m$ to the data in a window of width $w$, centered at each pixel. The $0^{\textrm{th}}$ order coefficient gives the value of the smoothed data at the central pixel, and successive coefficients yield the derivatives of the smoothed data.  This is a computationally efficient operation because the fit can be performed using linear algebra. The IDL library function \texttt{SAVGOL} returns the desired S-G kernel, but does not incorporate slope error estimates. We wrote our own S-G filter function, which smooths and differentiates the spectrum, and simultaneously returns slope error estimates derived from the input \apogee~ flux errors. The function performs the following calculation.

We describe the polynomial fit about a given pixel $\lambda_0$, as 
\begin{equation}
   Y(z) = a_0 + a_1\,z + \dots + a_m\,z^m
\end{equation}
where $z$ is a coordinate describing the distance from the central point $\lambda_0$, $z = \frac{\lambda-\lambda_0}{\Delta \lambda}$. The coefficients are found by solving

\begin{equation}
    \textbf{J}\bm{a} = \bm{y}
\end{equation}
where $\textbf{J}$ is a matrix whose columns are $\left[\bm{1}, \bm{z}, \bm{z}^2,\dots,\bm{z}^m\right]$, and $\bm{y}$ is the data in the window of width $w$, centered at $\lambda_0$.

The solution, given perfect data, is:

\begin{equation}
    \bm{a} = \left(\textbf{J}^T\textbf{J}\right)^{-1}\textbf{J}^T\bm{y} 
\end{equation}
We incorporate the flux errors through the covariance matrix in the following way:

\begin{equation}
    \bm{a} = \left(\textbf{J}^T\textbf{C}^{-1}\textbf{J}\right)^{-1}\textbf{J}^T\textbf{C}^{-1}\bm{y} \equiv \textbf{K} \bm{y}
\end{equation}
The result, $\textbf{K}$, is an $m$ degree by $w$ pixel matrix. The $j^{\textrm{th}}$ row of $\textbf{K}$ is a Savitsky-Golay convolution kernel corresponding to order $j$. The smoothed (and optionally differentiated) spectrum is given by the convolution of the kernel of desired order and the data, scaled by a normalization term for orders $m \ge 1$. 
That is, 

\begin{equation}
    \frac{\textrm{d}^mf}{\textrm{d}\lambda^m} = \frac{m!}{\Delta\lambda^m} \, \textbf{K}_m \ast I(\lambda)
\end{equation}
where $m$ is the desired order of differentiation and $\Delta\lambda$ is the wavelength spacing of the spectrum.  Specifically, our first derivative is given by: 

\begin{equation}
\frac{\textrm{d}f}{\textrm{d}\lambda } = \frac{1}{\Delta\lambda} \, \textbf{K}_1 \ast I(\lambda)
\end{equation}

The uncertainties of the filtered data of order $m$ at wavelength $\lambda$ are the diagonal elements of the coefficient covariance matrix:

\begin{equation}
    \Sigma_{\lambda} = \left(\textbf{J}^T\textbf{C}^{-1}\textbf{J}\right)^{-1}
\end{equation}
Therefore, the variance of the slope at pixel $\lambda$, $\sigma^{\prime\,2}_{\lambda}$, is element $[1,1]$ of $\Sigma_{\lambda}$. We compared the results of our custom S-G filter with those output by the IDL \texttt{SAVGOL} function, and found them to be in agreement. We tested our error estimates against those computed by two other methods. The first method was the straight-forward propagation of error, assuming a simple calculation of the slope $\df_\lambda = \frac{f_{\lambda, i+1} - f_{\lambda, i-1}}{2\Delta\lambda}$. The second method was the sampling of the flux errors through a Monte Carlo approach. For each spectrum, we generated $1\e3$ realizations with random fluctuations dictated by the \apogee~ flux error vectors. For each realization, we computed the slope at each pixel using the S-G filter, and took the standard deviation of the slopes as the slope error. All three methods showed broad agreement. 

We use the \apogee~ pixel masks (HDU 3 of the apStar files) to identify bad pixels in the apStar spectra, which are effectively masked out in the flux slope estimates. Like \citet{Cas16}, we use all mask bits other than $9$, $10$, and $11$, which correspond to persistence effects. We note that for the purpose of this step, we set the error in masked pixels to $\sigma_{\lambda\textrm{masked}} = 1$, and cap the error in unmasked pixels at the same value.  We also cap the flux in masked pixels at $1$. Although we are artificially reducing flux errors, we argue that all pixels in a continuum-normalized spectrum should, astrophysically speaking, have flux between $0$ and $1$ (modulo noise). Capping the flux error at $1$ allows us to compute slopes that are still extremely uncertain while avoiding floating point overflows. 

\section{KOI 3052}
\label{sec:outlier}
One star in our training sample appeared as an obvious outlier. For KOI 3052, the CKS pipeline measured $\vsini_{\textrm{CKS}} = \unit[3.4]{\kms}$, and we consistently measured $\vsini_{\textrm{MODEL}} \sim \unit[10]{\kms}$. While the CKS estimates are made from higher-resolution spectra, the ASPCAP estimate supported our finding with $\vsini_{\textrm{ASPCAP}} = \unit[9.7]{\kms}$. Upon further visual inspection, the spectra suggest that both estimates are correct. That is, the CKS spectrum appears to be relatively unbroadened, while the APOGEE spectrum appears to be noticeably broadened. The period of KOI $3052$ has been measured in several analyses of Kepler photometric data to be between $P = \unit[27.75]{days}$ and $P = \unit[30.17]{days}$ (\citealt{McQ13}, \citealt{Rein13}, \citealt{Wal13}, \citealt{Maz15}). Using the CKS estimate of $R_\ast = \unit[0.83]{R_\sun}$, the periods suggest an equatorial velocity of $v_{\textrm{eq}} \sim \unit[1.4]{\kms}$. Because of the projection effect, it must be the case that $\vsinie \le v_{\textrm{eq}}$. Although unlikely, it is possible that the measured rotation period is a harmonic of the `true' rotation period, as a period of $P \sim \unit[14]{days}$ would be within one sigma of the CKS \vsini. \nt{Of course, the CKS \vsinie may be overestimated, as it is close to the resolution-limited detection floor of the CKS spectra.} The much broader APOGEE spectrum may be the result of an unresolved binary companion. Because the analysis presented here relies upon the agreement between the labels and the data, and because the ASPCAP \vsinie appears consistent with the observed near-infrared spectrum for KOI 3052, we used the ASPCAP \vsinie value as the reference label for this star only.

\software{idlutils \url{http://www.sdss.org/dr13/software/idlutils/}, amoeba.pro}  

\clearpage

\clearpage

\clearpage

\clearpage

\clearpage

\clearpage

\clearpage

\clearpage

\clearpage      

\clearpage

\end{document}